\documentstyle[12pt,epsfig]{article}
\newcommand{\plb}[3]{Phys.~Lett.~B #1 (#2) #3}

  \newcommand{\ccaption}[2]{
    \begin{center}
    \parbox{0.85\textwidth}{
      \caption[#1]{\small{{#2}}}
      }
    \end{center}
    }
\begin{document}
\begin{titlepage}
\nopagebreak
{\begin{flushright}{
 \begin{minipage}{5cm}
	CERN-TH/2002-287 \\
	FNT/T-2002/16 \\
	hep-ph/0210261
\end{minipage}}\end{flushright}}
\vfill
\begin{center}
{\Large\bf $\mathbf{b\bar{b}}$ final states in
Higgs Production via \\ Weak Boson Fusion at
the LHC~\footnote{The work of MLM and FP is
        supported in part by the EU Fourth Framework Programme
        ``Training and Mobility of Researchers'', Network ``Quantum
        Chromodynamics and the Deep Structure of Elementary
        Particles'', contract FMRX--CT98--0194 (DG 12 -- MIHT). MM and
        RP acknowledge the financial support of the European Union
        under contract HPRN-CT-2000-00149. 
        RP thanks the finantial support of MIUR under 
        contract 2001023713-006. ADP is supported by a
      M. Curie fellowship, contract HPMF-CT-2001-01178.}}
\end{center}
\vfill
\begin{center}
{\large
M.L.~Mangano$^{(a)}$, M.~Moretti$^{(b)}$, F.~Piccinini$^{(a)}$\footnote{On 
leave of absence from INFN Sezione di Pavia, Italy.}, 
R.~Pittau$^{(c)}$ and A.D.~Polosa$^{(a)}$}
\end{center}
\vskip 0.5cm
\begin{center}
$^{(a)}$ CERN Theory Division, 1211 Geneva 23, Switzerland\\
\vskip 12pt\noindent
$^{(b)}$ Dipartimento di Fisica - Universit\`a di    Ferrara, and \\
INFN - Sezione di Ferrara, Ferrara, Italy
\vskip 12pt\noindent
$^{(c)}$ Dipartimento di Fisica - Universit\`a di Torino, and \\
INFN - Sezione di Torino, Torino, Italy
\end{center}
\vfill
\begin{abstract}
We present a study of the Higgs production at the LHC via Weak Boson
Fusion, with the Higgs boson decaying into a $b\bar{b}$ pair.  A detailed
partonic LO calculation of all the potential backgrounds is
performed. We conclude that this channel for Higgs production can be
extracted from the backgrounds, and present our estimates of the
accuracy in the determination of the $Hb\bar{b}$ Yukawa coupling.

\vskip 1cm
\end{abstract}
CERN-TH/2002-287\hfill \\
Jan 10, 2003 \hfill  
\vfill       
\end{titlepage}

%%%%%%%%%%%%%%%%%%%%%%%%%%%%%%%%%%%%%%%%%%%%%%%%%%%%%%%%%%%%%%%%%%
\newpage

\section{Introduction}
\label{intro}
A Higgs boson in the so-called low-mass region ($115<m_H({\rm GeV})
<140$) decays predominantly in $b\bar b$ final states.
Due to the large inclusive QCD backgrounds, detection of this decay is
however extremely challenging. In particular, the extraction of the
most copious signal, namely inclusive $gg\to H \to b\bar{b}$
production, has never been shown to be viable. 
The only production channels which have so far been proven to be
suitable for a determination of the $Hb\bar{b}$ coupling are
the associate production $H t \bar t$ and $HW$~\cite{cmsnote,
cmsnote2}. 

In this note we document a study of the $H\to b\bar{b}$ decay in
the electroweak boson fusion (WBF)
production channel and of its backgrounds, and we discuss the potential
of this process  for the determination of the $y_{Hbb}$ Yukawa coupling. 
The signal rate is proportional to the product of the 
$y_{HVV}$ coupling, where $V$ denotes 
a weak $W$ or $Z$ boson, times the ${\cal B}(H\to b\bar{b})$ 
branching ratio. The contamination to the signal coming from
QCD production of  Higgs plus two jets (mediated by a loop of virtual
top quarks) are not included in this analysis. Following the study of
ref.~\cite{duca},  these will be 
suppressed by the particular set of kinematical cuts 
chosen in our analysis (see Section 2).

The results obtained are based on a leading order 
partonic calculation of the matrix elements (ME) describing signal and 
background processes. The latter include the following channels: 
QCD $b \bar b j j$ production, 
$Z(\to b \bar b) j j$, $W/Z(\to j j) b \bar b$, 
$t \bar t \to b \bar b + jets$,  QCD four jets production (where two
light jets are misidentified as generated by $b$ quarks), and
contributions from multiple overlapping events.

We identify a set of  kinematical cuts leading to signal significances
in the range of $2-5\sigma$, depending on the Higgs mass.
In the lowest mass region, this provides a determination 
of the ${\cal B}(H\to b \bar{b})$
branching ratio with a precision of the order of $20\% $. 
The $H\to b\bar{b}$ decay in the WBF channel could be used 
together with other processes already examined in literature for a model
independent determination of the ratio of Yukawa couplings
$y_{Hbb}/y_{H\tau\tau}$~\cite{zepp}.
We therefore conclude that the $H\to b\bar{b}$ channel produced in association 
with two jets should be considered as an additional channel to be 
exploited for interesting measurements of the Higgs couplings to fermions.

This letter is organized as follows. In Section 2 we describe 
the kinematical constraints introduced to perform the event selection. 
Section 3 is devoted to the discussion of signal and backgrounds,
while  the signal significance and the accuracy of the branching ratio 
$H\to b\bar{b}$ and Yukawa coupling determination are presented in Section 4. 
In the Conclusions we summarise and discuss our final results. 

\section{Event selection}
The choice of selection criteria is guided by two main requirements:
the optimization of the signal significance ($S/\sqrt{B}$), and the
compatibility with trigger and data acquisition constraints.
The main features of the signal, to be exploited in the event
selection, are: presence of two, high-$p_{\rm T}$, $b$  jets,
 showing an invariant-mass
peak; presence of a pair of jets in the forward and backward rapidity
regions. 
In principle such a signal could also exhibit rapidity gaps, due to
the colour-singlet exchange of EW bosons among the incoming
hadrons; this fact has been used recently in~\cite{DeRoeck:2002hk}.
Because of the high luminosity (and the large number of
overlapping events) required to study this final state,
and because of the large emission rate for extra jets in
WBF processes (see~\cite{alpgen}), we do not feel comfortable with
applying this additional constraint in our study.

\begin{figure}
\begin{center}
\epsfig{file=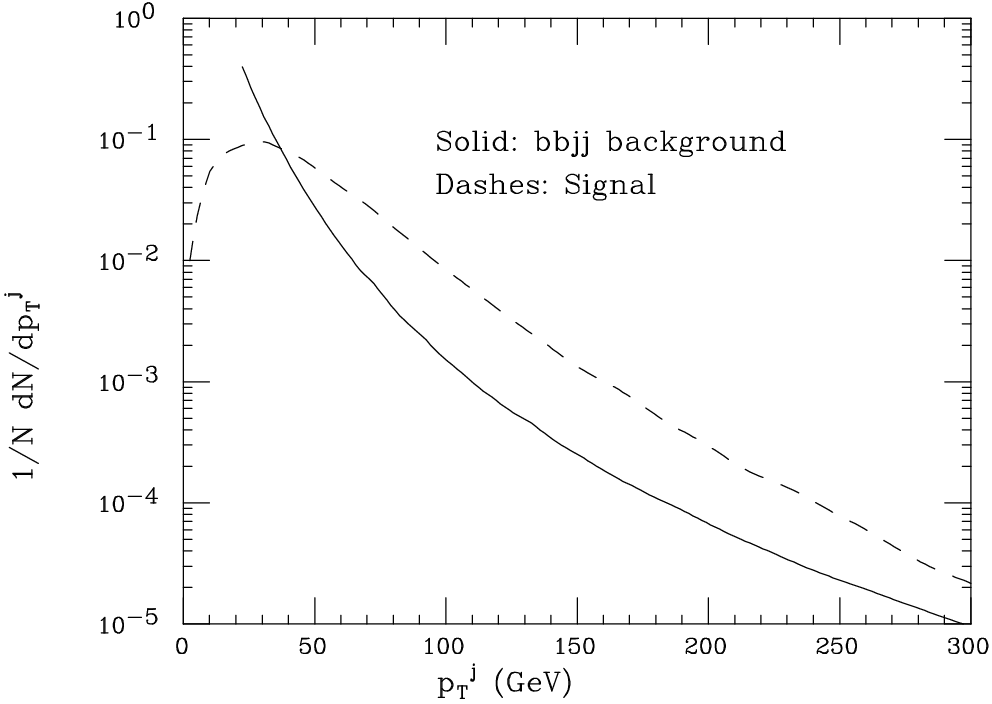,width=0.7\textwidth}
\vskip -2mm
\ccaption{*}{The $p_{\rm T}^j$ distributions are shown: high 
$p_{\rm T}^j$ regions are more 
suppressed in the $b\bar{b}jj$ QCD background (solid) 
with respect to the signal (dashes). The inclusive 
distributions shown are normalised to the same cross section.
\label{fig:ptj}
}
\end{center}
\end{figure}

The  tagging of the $b$ jets is only possible in the central region
$\vert \eta_b \vert<2.5$. The efficiency of the tagging algorithm,
furthermore, suggests using a $p_{\rm T}^j$ cut as
large as possible. Since the measurement of the Higgs boson in this
channel will  take place only after its discovery and the
determination of its mass, we can optimize the mass requirement by
selecting only $b$ pairs in a mass window centred around the known
value of $m_H$, up to the dijet mass resolution.
These considerations lead to the following set of cuts:
\begin{eqnarray}
p_{\rm T}^{b} &>& 30~{\rm GeV} \label{eq:bcuts1}\\
|\eta_{b}| &<& 2.5 \label{eq:bcuts2}\\
\Delta R_{bb}&>&0.7 \label{eq:bcuts3}\\
|m_{bb}-m_H |&<& \delta_m\cdot m_H,\  \label{eq:bcuts4}
\end{eqnarray}
$\delta_m$ being the experimental resolution~$\simeq 12\%$.
Given the very small width of the Higgs boson in the mass range we
shall consider ($m_H<140$~GeV), this last requirement  reduces the
 signal to  68\% of what obtained with perfect mass
resolution. 
In the following we shall assume a $b$-tagging efficiency $\epsilon_b=0.5$.
While harder cuts on $p_{\rm T}^b$ would improve the $S/B$ ratio, they
would also risk sculpting the mass distribution, setting a higher
value for the dijet mass threshold and therefore making it harder to
extract the background shape directly from the data.

\begin{figure}
\begin{center}
\epsfig{file=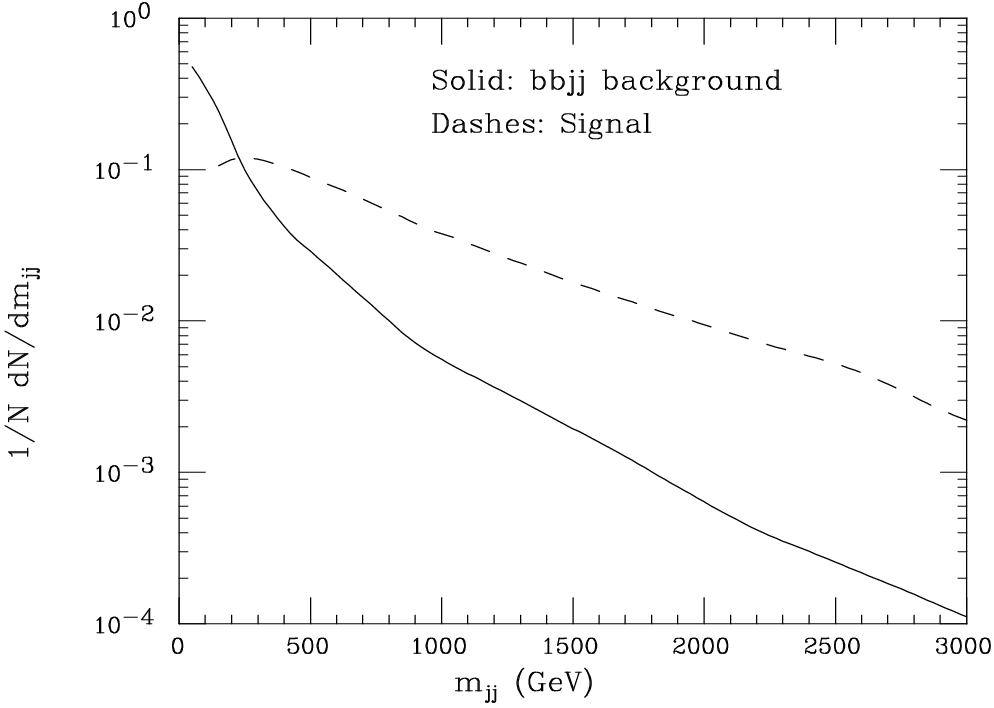,width=0.7\textwidth}
\vskip -2mm
\ccaption{*}{The distribution for $m_{jj}$ is shown 
both for the signal (dashes) and for the $b\bar{b}jj$ QCD background
(solid). The inclusive distributions shown are 
normalised to the same cross section.
\label{fig:mjj}
}
\end{center}
\end{figure}

The large momentum exchange required for the emission of the
space-like gauge bosons will lead to a hard $p_{\rm T}^j$ spectrum for the
forward and backward light 
jets. This is clearly shown in Fig.~\ref{fig:ptj}\footnote{The distributions shown in
the first two figures are obtained by applying  no cuts to the signal,
 and  the following minimal cuts on the background: $p_{\rm T}^j>20$~GeV,
$|\eta|<5$~GeV, $\Delta R_{jj,bb,jb}>0.2$.}, 
where we see that the jet $p_{\rm T}$
peaks at approximately 30~GeV. The spectrum of typical QCD backgrounds
will viceversa peak at low $p_{\rm T}^j$.  
The large momentum of the forward jets, and their large rapidity
separation, favours large dijet invariant masses, as can be seen from
Fig.~\ref{fig:mjj}. The cuts we select for the two jets are:
\begin{eqnarray}
p_{\rm T}^{j} &>& 60 \;\rm{or}\; 80~{\rm GeV} \label{eq:jcuts1}\\
|\eta_{j_1}-\eta_{j_2}|&>&4.2  \label{eq:jcuts2}\\
\Delta R_{jj} ,\Delta R_{jb}&>&0.7  \label{eq:jcuts3}\\
m_{jj}&>&1000~{\rm GeV}.  \label{eq:jcuts4}
\end{eqnarray}
The large $p_{\rm T}^j$ cut is driven by the requirement that
trigger rates be kept at acceptable levels (see later). We present the
two cases of 60 and 80~GeV to display the sensitivity to this
threshold. A final choice will presumably only be possible with a
complete detector simulation, or once the background data will be available.
As we will comment later, the cut on $p_{\rm T}^j$ above $80$~GeV is 
also very efficient in decreasing the backgrounds due to
multiple overlapping events. 
The large mass cut is selected to reduce as much as possible the QCD jet
backgrounds. This cut, in addition to the rapidity cut,
is also efficient in removing the contamination
from the process $gg\to H gg$, as shown in ref.~\cite{duca}. 

In addition to the above cuts, we shall consider two alternative
selection criteria for the light-jet rapidities, labelled $(a)$ and
 $(b)$.  The case $(a)$ is given by:
\begin{equation}
2.5  <|\eta_{j}| < 5,\;\;\; \eta_{j_1}\eta_{j_2}<0,\label{eq:etacuta}
\end{equation}
while for the the case $(b)$, we only have the condition:
\begin{equation}
|\eta_{j}| < 5.\label{eq:etacutb}
\end{equation}
In the case $(b)$ we verified that requiring
$m_{jj}>1000$~GeV  forces the product $\eta_1\cdot
\eta_2$ to be negative for the largest fraction of the events.

By inspection of the differential distributions for the 
variable $\Delta R_{bb}$ we find that cutting $\Delta R_{bb}<2$ 
for the configuration $(a)$ gives an additional enhancement 
of the signal with respect to the backgrounds.

\section{The study of signal and backgrounds}
The  background sources we considered include:
\begin{enumerate}
\item   QCD production of $b\bar{b}jj$ final states, where $j$ indicates a jet originating from a
light quark ($u,d,s,c$) or a gluon;
\item QCD production of  $jjjj$ final states. 
\item Associated production of $Z^*/\gamma^* \to b\bar{b}$ and light
jets, where the invariant mass of the $b\bar{b}$ pair is in the Higgs
signal region either because of imperfect mass resolution, or because
of the high-mass tail of the intermediate vector boson.
\item $t\bar{t}$ production
\item $t\bar{t}j$ production
\item $b\bar{b}jj$ and $jjjj$  production via overlapping events. 
\end{enumerate}
The cases with 4 light-jet events are considered since the experimental
resolution leads, for any tagging algorithm, to a finite probability
of $b$ tags in light jets ({\em fake tags}). We shall label light jets
mistagged as $b$ jets with the notation $j_b$, and  assume two
possible values of fake tagging efficiencies $\epsilon_{fake}$, 1\%
and 5\%. While the first choice is probably optimistic, given the
presence of real secondary vertices in jets containing a charm quark,
the second is likely to be too conservative. As we shall see, however,
the requirement of tagging both $b$ jets renders in any case the
backgrounds with real $b$ quarks the dominant ones.

The calculation of signal and background events is based on the
numerical iterative procedure ALPHA~\cite{alpha}, as implemented in the 
library of MC codes ALPGEN~\cite{alpgen}.  While ALPGEN allows
for the full showering of the final states, both in the case of
signals and backgrounds, all our calculations are limited to the
parton level. This is because a realistic estimate of the rates would
anyway require a full detector simulation, which is beyond the scope
of this paper.

%%%%%%%%%%%%%%%%%%%%%%% I TABLE
{\renewcommand{\arraystretch}{1.2}
\begin{table}[h]
\begin{center}
\ccaption{*}{Signal and background events for configuration (a), with
$p_{\rm T}^j>60$~GeV, 
for three possible values of the Higgs mass. 
 $Q^2=\langle {p_{\rm T}}^2 \rangle$.  
The $jjjj$ entry  includes the
squared $b-$mistagging efficiency ($\epsilon_{fake} = 0.01$). 
The first raw relative to the $Z^*/\gamma^*$ contribution refers to the
effect of the physical mass tail, while the second raw refers to the
finite experimental $Z$ mass 
resolution, $(\delta m_Z/m_Z=0.12)$.
The integrated luminosity is 600~fb$^{-1}$.
The PDF set used is CTEQ4L. See the text for the description of other,
smaller, backgrounds.
\label{tab:sb60a}
}
\vskip 2mm
\begin{tabular}{||l|l|l|l||}\hline
$m_H$ & 115~GeV  & 120~GeV  & 140~GeV \\
\hline
Signal  & $3.0\times 10^3$  & $2.8\times 10^3$  & $1.1\times 10^3$ \\
\hline
$b \bar b j j $ & $8.6\times 10^5$  & $8.0\times 10^5$   & $5.7\times 10^5$ \\
\hline
$j_b j_b j j$ & $6.4\times 10^3$ & $6.1\times 10^3$ & $4.1\times 10^3$\\
\hline
$(Z^*/\gamma^*\to b\bar{b}) jj$ & $5.5\times 10^2$ & $3.8\times 10^2$ & 
$1.0\times 10^2$  \\
\hline
$ (Z\to b\bar{b})_{\rm res}jj  $ & $1.3\times 10^3$ & $6.8\times 10^2$
& $1.1\times 10^1$ \\ 
\hline
$j_b j \oplus j_b j$  & $7.5\times 10^3$  & $7.9\times 10^3$  & $9.0\times 10^3$  \\
\hline
\end{tabular}            
\end{center}
\end{table}
}

%%%%%%%%%%%%%%%%%%% II TABLE
{\renewcommand{\arraystretch}{1.2}
\begin{table}[h]
\begin{center}
\ccaption{*}{Same as Table~\ref{tab:sb60a}, for configuration (b).
\label{tab:sb60b}
}
\vskip 2mm
\begin{tabular}{||l|l|l|l||}\hline
$m_H$ & 115~GeV  & 120~GeV  & 140~GeV \\
\hline
Signal  & $1.3\times 10^4$ & $1.2\times 10^4$ &  $6.2\times 10^3$ \\
\hline
$b \bar b j j $ & $6.0\times 10^6$  & $5.3\times 10^6$   & $4.7\times 10^6$ \\
\hline
$j_b j_b j j$   & $1.2\times 10^5$  & $1.1\times 10^5$  & $1.1\times 10^5$ \\
\hline
$(Z^*/\gamma^*\to b\bar{b}) jj$ & $4.5\times 10^3$ & $2.8\times 10^3$ & 
$1.1\times 10^3$ \\
\hline
$ (Z\to b\bar{b})_{\rm res}jj  $  & $1.6\times 10^4$ & $8.3\times 10^3$ & $7.7\times 10^2$ \\
\hline
$j_b j \oplus j_b j$  &  $1.8\times 10^4$ &$1.9\times 10^4$ &$2.3\times 10^4$\\
\hline
\end{tabular}            
\end{center}
\end{table}
}
%%%%%%%%%%%%%%%%%%%%%%% I TABLE 80 ptjmin
{\renewcommand{\arraystretch}{1.2}
\begin{table}[h]
\begin{center}
\ccaption{*}{Same as Table~\ref{tab:sb60a},  with
$p_{\rm T}^j>80$~GeV.
\label{tab:sb80a}
}
\vskip 2mm
\begin{tabular}{||l|l|l|l||}\hline
$m_H$ & 115~GeV  & 120~GeV  & 140~GeV \\
\hline
Signal  & $1.3\times 10^3$  & $1.2\times 10^3$ & $5.2\times 10^2$ \\
\hline
$b \bar b j j $ & $2.4\times 10^5$  & $2.3\times 10^5$   & $1.9\times 10^5$ \\
\hline
$j_b j_b j j$ & $2.6\times 10^3$ & $2.3\times 10^3$ & $1.8\times 10^3$\\
\hline
$(Z^*/\gamma^*\to b\bar{b})jj  $ & $1.1\times 10^2$ & $6.6\times 10^1$ & 
$1.3\times 10^1$  \\
\hline
$ (Z\to b\bar{b})_{\rm res}jj  $  & $6.2\times 10^2$ & $3.4\times 10^2$ & $0.5\times 10^1$ \\
\hline
$j_b j \oplus j_b j$  & 
$2.9\times 10^2$  & $3.2\times 10^2$  & $4.5\times 10^2$  \\
\hline
\end{tabular}            
\end{center}
\end{table}
}
%%%%%%%%%%%%%%%%%%% II TABLE PTJ > 80
{\renewcommand{\arraystretch}{1.2}
\begin{table}[h]
\begin{center}
\ccaption{*}{Same as Table~\ref{tab:sb80a}, for configuration (b).
\label{tab:sb80b}
}
\vskip 2mm
\begin{tabular}{||l|l|l|l||}\hline
$m_H$ & 115~GeV  & 120~GeV  & 140~GeV \\
\hline
Signal  & $6.5\times 10^3$ & $6.4\times 10^3$ &  $3.1\times 10^3$ \\
\hline
$b \bar b j j $ & $2.8\times 10^6$  & $2.2\times 10^6$   & $2.1\times 10^6$ \\
\hline
$j_b j_b j j$ & $5.6\times 10^4$  & $5.3\times 10^4$  & $5.2\times 10^4$ \\
\hline
$(Z^*/\gamma^*\to b\bar{b})jj$ & $3.0\times 10^3$ & $1.9\times 10^3$ & 
$7.5\times 10^2$ \\
\hline
$ (Z\to b\bar{b})_{\rm res}jj$ & $1.1\times 10^4$ & $6.0\times 10^3$ & $5.6\times 10^2$ \\
\hline
$j_b j \oplus j_b j$ & $1.1\times 10^4$& $1.2\times 10^4$ &$1.6\times 10^4$ \\
\hline
\end{tabular}            
\end{center}
\end{table}
}
The event rates are obtained using the parametrization of parton
densities  CTEQ4L. Given the overall uncertainties of the background
estimates, the results are not sensitive to this choice.  The
renormalization and factorization scales have been chosen equal
($Q$). In order to be conservative in the background estimates, we
selected as a default for our study a rather low scale, namely
$Q^2=\langle p_{\rm T}^2 \rangle$, where the average is taken over all
light and $b$ jets in the event\footnote{We also repeated our analyses
with $Q^2=m_H^2$, finding comparable results.}.  In view of the large
$\hat{s}$ values of the elementary processes involved, due in particular
to the large mass threshold for the pair of forward jets, we
believe that our background rates may be overestimated by a factor of
at least 2. In spite of this we prefered the conservative approach, in
order to present a worse-case scenario.
The backgrounds are much more sensitive to the scale choice than the
signal, due to the larger power of $\alpha_s$. The background
uncertainty will not however be a limitation to the experimental
search, since the background rate should be determined directly from the
data, as we shall discuss.

Tables~\ref{tab:sb60a}-\ref{tab:sb80b} present our results 
for signal and backgrounds, for
the following cases: ({\em{i}}) $p_{\rm T}^j>60$~GeV and rapidity
configuration (a); ({\em{ii}}) $p_{\rm T}^j>60$~GeV and rapidity
configuration (b); ({\em{iii}}) $p_{\rm T}^j>80$~GeV and rapidity
configuration (a); ({\em{iv}}) $p_{\rm T}^j>80$~GeV and rapidity
configuration (b). The numbers correspond to 
 $600$~fb$^{-1}$ of integrated luminosity, namely 
 the expected value for three years of running of ATLAS and CMS
 with an instantaneous luminosity of 
$10^{34}{\rm cm}^{-2}{\rm sec}^{-1}$. The numbers relative to final
states with mistagged jets include the square of the mistagging
probability $\epsilon_{fake}=0.01$.

We shall now discuss each individual background contribution in detail.

\subsection{Single-interaction events}
The 4-jet backgrounds originating from a single hard collision are
shown in the second and third rows of Tables~\ref{tab:sb60a}-\ref{tab:sb80b}. 
In the case of the $j_b j_b jj$ background, we accept all events in which at
least one pair of light jets passes the cuts in
eqs.(\ref{eq:bcuts1})-(\ref{eq:bcuts4}), and the other two jets satisfy
eqs.(\ref{eq:jcuts1})-(\ref{eq:jcuts4}), in addition to the
appropriate rapidity cut (eq.(\ref{eq:etacuta}) or
(\ref{eq:etacutb})). As anticipated, the contribution from real $b$
jets is the dominant one, even assuming $\epsilon_{fake}=0.05$. 

\begin{figure}
\begin{center}
\epsfig{file=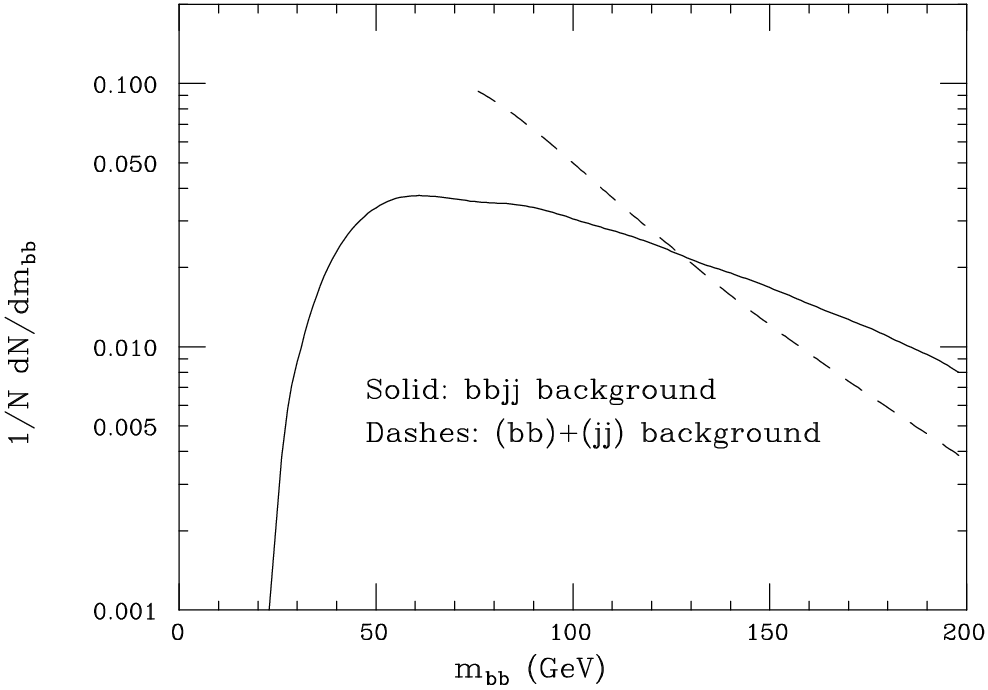,width=0.7\textwidth}
\vskip -2mm
\ccaption{*}{The distribution of the invariant mass of 
the system $b\bar{b}$ in the $b\bar{b}jj$ QCD background (solid line),
and in overlapping events of the type 
$(b \bar b) \oplus (j j)$ (dashed line). 
The curves are  normalised to the same cross section. \label{fig:mbb}
}
\end{center}
\end{figure}

From the numbers in the Tables~\ref{tab:sens60} and \ref{tab:sens80}, 
we see that the $S/\sqrt{B}$ can be as
large as $5$. However, the ratio $S/B$ is only a fraction of a
percent. This implies that the background itself will have to be known
with accuracies at the permille level.  There is no way that this
precision can be obtained from theoretical calculations. The
background should therefore be determined entirely from the data. We
expect our kinematical thresholds to be low enough not to sculpt the
shape of the $b\bar{b}$ mass distribution at masses close to the Higgs
mass. This is true for the leading 4 jet backgrounds, as shown in
Fig.~\ref{fig:mbb}.  The $b\bar{b}$ invariant mass of the simulated
$b\bar{b}jj$ background is shown here to be well behaved in the
$[100,150]$~GeV region. The distribution in the case of the $j_bj_bjj$
final states is similar.  As a result, we expect that the sidebands of
the Higgs signal (the regions of mass below $m_H(1 - \delta_m)$ and
above $m_H(1 + \delta_m)$) can be safely interpolated in the region
under the Higgs peak, similarly to what was done by UA2 in the
extraction of the $W/Z\to jj$ decay~\cite{Alitti:1990kw}.

For this extraction to be possible, however, full background samples
have to be collected. The large rate of untagged $jjjj$ events could
therefore give problems with the triggers and with the data
acquisition. This is because the $b$ tagging algorithm is typically
applied only offline, and therefore a number of untagged $jjjj$ events
larger than what is acceptable by the trigger and by the data
acquisition would force higher cuts, or a trigger prescaling, strongly
reducing the number of recorded signal events.  Removing the
fake-tagging probability from the numbers in the 
Tables~\ref{tab:sb60a}-\ref{tab:sb80b}, 
leaves untagged $jjjj$ rates in the range of few$\times 10^7$ and 
$10^9$, depending on whether configuration (a) or (b) is chosen. Since
the mass window for the signal is approximately 30~GeV wide, these
rates must be increased by a factor of 3-4, to allow for a sufficient
coverage of the sidebands of the $b\bar{b}$ mass distribution,
coverage which is required to enable the interpolation of the
background rate under the Higgs mass peak. The numbers in the 
Tables~\ref{tab:sb60a}-\ref{tab:sb80b} 
refer to 6 years of data taking, corresponding to $6\times 10^7$s,
distributed among the two experiments.  The result is a rate of events
to tape in the range of 1~Hz (for configuration (a) with 80~GeV jet
threshold) up to 50~Hz (for configuration (b) with 60~GeV jet
threshold). While a 1~Hz rate to tape is acceptable, 50~Hz would almost
saturate the expected data acquisition capability of 100~Hz.  In this
last case, some extra information would have to be brought into the
trigger. The best candidate is some crude $b$-tagging. If a rejection
against non-$b$ jets at the level of 20\% per jet could be achieved at
the trigger level, the rates would be reduced by a factor of 20, down
to perfectly acceptable levels.

While the above processes represent the largest contribution to the
backgrounds, the smoothness of their mass distribution in the signal
region allows to estimate their size with statistical accuracy,
without significant systematic uncertainties. The situation is
potentially different in the case of the backgrounds from the tails of
the $Z$ decays. The $Z$ mass peak is sufficiently close to $m_H$,
especially in the case of the lowest masses allowed by current limits,
to possibly distort the $m_{bb}$ spectrum and spoil the ability to
accurately reconstruct the noise level from the data.  The size of the
two possible effects (smearing induced by the finite experimental
energy resolution and the intrinsic tail of the Drell-Yan spectrum)
are given in the 4th and 5th rows of the
Tables~\ref{tab:sb60a}-\ref{tab:sb80b}.  Aside from the case of the
largest $m_H$ value, where these backgrounds are anyway negligible,
the dominant effect is given by the detector resolution.  For the
configurations (a) these backgrounds represent a fraction of the order
of at most 40\% of the signal, at small $m_H$, rapidly decreasing at
higher $m_H$. For the configuration (b), the rates are comparable to
the signal at low $m_H$.  A 10\% determination of these final states,
which should be easily achievable using the $(Z\to \ell^+\ell^-)jj$
control sample and folding in the detector energy resolution for jets,
should therefore be sufficient to fix these background levels with the
required accuracy.  As for the contribution of the on-peak $(Z\to
b\bar{b})jj$ events to the determination of the sideband rates, we
verified that their impact is negligible. We obtain a number of the order of
60K events with 600~fb$^{-1}$ in the mass range 83-100 GeV, for
configuration (b) and $p_T>80$~GeV for the forward jets. These events
can therefore be subtracted from the sidebands with a statistical
accuracy better than 1\% using the measurement of the on-peak $(Z\to
\ell^+\ell^-)jj$ final states. It should be pointed out that
extrapolating from the leptonic to the $b\bar{b}$ rates with this
accuracy requires a matching precision in the knowledge of the tagging
efficiencies, something which remains to be proven.

Before concluding the list
of single-interaction backgrounds, we briefly comment on the smaller
contributions, $pp\to t\bar{t}$ and $pp\to t\bar{t} j$, with $t$
decaying hadronically.  Before applying the cuts, we adopt a
clustering algorithm for the jets coming from the decay of a $W$. We
sum the four-momenta every time the separation between the two jets is
below the threshold $\Delta R = 0.4$.  This happens quite often, since
in order to have a pair of jets in the event with an invariant mass
above 1~TeV at least one of the two $W$s coming from the $t$ decays
must have a large boost. After this clustering algorithm, using the
event selection $(b)$, about $300$ $t \bar t j$ events survive the
cuts at $600$~fb$^{-1}$, while the number of $t \bar t $ events is
negligible.  The configuration $(a)$ leads to even smaller rates.  The
absolute rate can be fixed using the data, by reconstructing the
individual tops. This should be particularly simple, since the request
of large dijet mass forces the $t$ and $\bar{t}$ to be very well
separated, and the large momentum of the $W$'s will reduce the
combinatorial background in the association of the $b$ jets with the
$W$ jets.

\subsection{Overlapping events}
We come now to the study of events due to the superposition of
multiple $pp$ interactions.  The reason why these events are a
potential problem is that while production of large dijet invariant
masses in individual events is strongly suppressed energetically,
these can accidentally appear when mixing jets produced in separate
events (after all the overall energy available in 2 collisions is
twice that for a single $pp$ collisions): for example, we can consider
two events, one in which a small-mass dijet pair is produced with
large positive rapidity, the other in which a low-mass pair is
produced at large negative rapidity; the pairing of jets from the two
events will lead to large rapidity separations, and to large dijet
masses.

In the simplest case of two overlapping events, we have four possible
combinations of events leading to a $b\bar{b}jj$ background:
$(jj)~\oplus~( b \bar{b})$, $(jj)~\oplus~(j_b j_b)$, $(jj_b)~\oplus~(
j j_b)$ and $(b \bar{b})~\oplus~( b \bar{b})$, where $(ab)\equiv pp\to
ab$.  Since we do not veto on the presence of extra jets, triple
events such as $(j_1j_b)~\oplus~ (jj_2)~\oplus~( j_bj)$ are also
possible.  The probability of having $n$ simultaneous events with a
$jj$ final state during a bunch crossing, assuming a bunch crossing
frequency of $(25$ ns)$^{-1}$, is given by the Poisson probability
distribution function $\pi_n(\mu)$ with average $\mu = 0.25
\times \sigma (pp\to jj)/{\rm mbarn} \times {\cal L}/{\cal
L}_0$, where ${\cal L}$ is the instantaneous luminosity and ${\cal
L}_0 = 10^{34}{\rm cm}^{-2}{\rm sec}^{-1}$.

To estimate the rates, we first generate a sample of unweighted events
of the type $pp\to jj$. We then randomly extract from this sample
$n$-tuples of dijet events, which are associated to events where $n$
dijet pairs from $n$ proton-proton collisions are created in the same
bunch crossing. The background can be then estimated as:
\begin{equation}
N_{bg} \; = \; B\times(\pi_2(\mu)p_2+\pi_3(\mu)p_3+...),
\end{equation}
where B is the number of bunch crossings accumulated during the run
time, and $p_n=f_n/N_n,~(n=2,3)$, where $N_n$ is the total
number of $n$-tuple events generated, 
$f_2,f_3$ are the number of double and triple events 
passing the selection cuts found in the sample of generated events.
Ellipses denote simultaneous collisions of higher order.
Since $\pi_n(\mu)$ drops quite rapidly with increasing $n$, we limit 
our analysis at $n=3$. The above formula can be easily modified to
include the presence of $\sigma(pp\to b\bar{b})$ events.
All numbers given below refer to the case of high luminosity, namely 
 $10^{34}$cm$^{-2}$s$^{-1}$. Since these rates scale quadratically,
they should be reduced by a factor of 100 in the case of 
$10^{33}$cm$^{-2}$s$^{-1}$.

\begin{figure}
\begin{center}
\epsfig{file=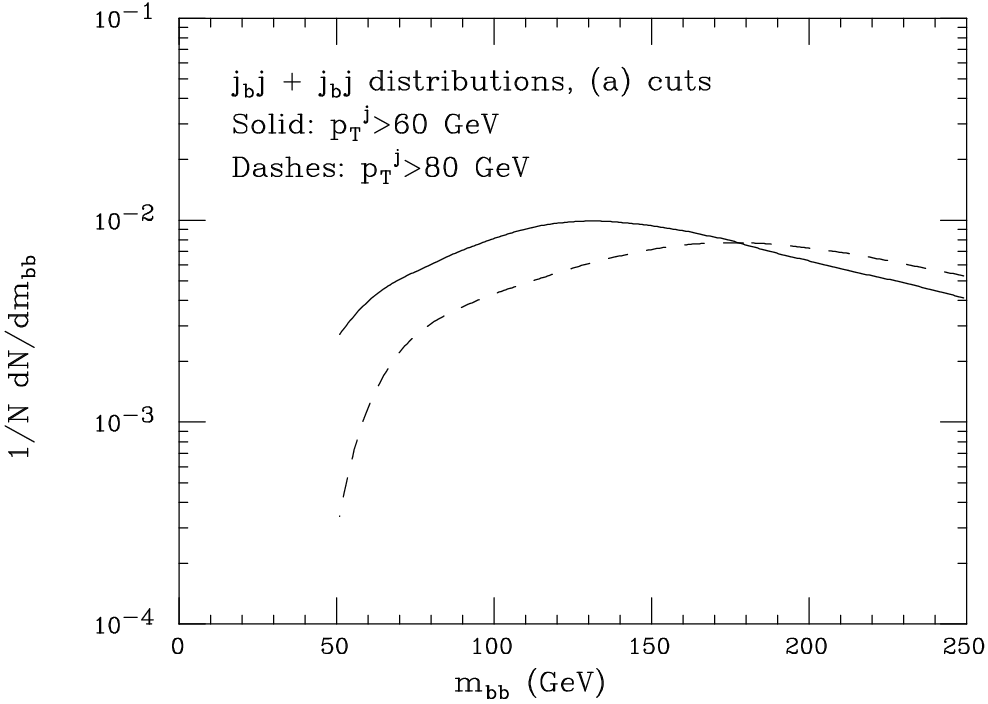,width=0.7\textwidth}
\vskip -2mm
\ccaption{*}{The distribution of the invariant mass of 
the system $b\bar{b}$ in the $j_b j \oplus j_b j$ multiple-collision 
QCD background, for configuration (a). 
\label{fig:mjbjb}
}
\end{center}
\end{figure}
We verified that the most dangerous background comes from events of
the type $(jj_b) \oplus (jj_b)$. The main reason is as follows: since
the forward, non-tagged jets are required to have a large $p_{\rm T}$
threshold (60 or 80~GeV), the fake $b$ jets in the central region will
inherit the same transverse momentum cut, as they are produced
back-to-back with the related forward jet. As a result, the invariant
mass spectrum of the $j_bj_b$ pair will have a shape peaked at about
twice the cut, and therefore right in the middle of the signal
region. Typical shapes of the $m_{bb}$ spectra are given in
Fig.~\ref{fig:mjbjb}, for configuration (a) (The shapes for
configuration (b) are very similar). 
In the case of 60~GeV, the signal regions are
right in the middle of the background peak, or on its rising slope; 
this makes the
background estimate very sensitive to the assumed energy resolution,
both in the forward region (since the energy scale in the forward
region affects the onset of the trigger for the forward jets, thus
affecting the spectra of the central jets recoiling against them) and
in the central region as well (since the mass spectrum is rapidly
rising in the 100-150~GeV range. Our results were obtained by assuming
a forward jet energy resolution given by $\sigma_{fwd}=\sqrt{E} \
\oplus 0.07\, E$, in addition to the 12\% mass resolution used earlier for
the central jets. The distributions in Fig.~\ref{fig:mjbjb} include
this resolution smearing. The rates obtained after including the resolution
effects are approximately twice as large as those obtained with
perfect resolution, stressing the importance of these effects.
In absolute terms, the Tables~\ref{tab:sb60a}-\ref{tab:sb80b} 
show that these contributions are of the same order of
magnitude as the signal when $p_{\rm T}^j>60$~GeV is used, but much
smaller when the higher $p_{\rm T}^j$ thereshold is used. In the former
case, these final states are a
potential threat, unless a way can be found to estimate from the data
their exact size. This cannot be done using the mass spectrum in the
sideband regions, since the rate is too small compared to the leading
4-jet processes. We believe that it should be possible however to use
the distribution of the $z$ vertex separation between the two events
as a diagnostic tool. Since the two tagged jets come from different
$pp$ events, and given that the spread of the interaction point in $z$
is of the order of few cm, the fraction of overlapping events where
the $z$ positions of the two vertices cannot be separated should be of
the order of 10\%, a number measurable by
extrapolating the $\Delta z$ distribution from large values, down to
the range in which $\Delta z$ is of the order of the experimental
resolution.

Other sources of backgrounds from overlapping events are less
dangerous. Events where the $b \bar{b}$ or $j_b j_b$ pair comes from
the same hard interaction ($(b\bar{b})\oplus (jj)$ and $(j_bj_b)
\oplus (jj)$)
have a smooth mass spectrum in the
100-150~GeV region, and rates smaller than
those of the
single-interaction $b\bar{b}jj$ or $j_bj_b jj$ events. The mass
spectrum of $(b\bar{b})\oplus (jj)$ events  is shown in
Fig.~\ref{fig:mbb}\footnote{The sharp threshold at approximately 70~GeV is due
to the fact that the $b$ and $\bar{b}$ are mostly produced
back-to-back, coming from a $2\to 2$ scattering; in the case of the
single-interaction $b\bar{b}jj$ events the $b$ and $\bar{b}$ can be
produced at relative angles as small as allowed by the $\Delta R_{bb}>0.7$ cut,
and the threshold onset is smoother.}.
Their contribution can
therefore be estimated precisely from the data\footnote{Of course
their individual contribution may not be easily obtained; what can be
estimated is the overall rate of 4-jet events, including both double-
and single-collision contributions.}. In the specific case of
$m_H=120$~GeV, for example, we obtain the following numbers of events:
$10^5$ and $4\times 10^5$ 
$ (jj)~\oplus~ (b\bar{b})$ events 
for $p_{\rm T}^j>60$~GeV in the  configurations (a) and (b), respectively;
 $6 \times 10^4$ and $2\times 10^5$ 
$ (jj)~\oplus~ (b\bar{b})$ events 
for $p_{\rm T}^j>80$~GeV in the  configurations (a) and (b), respectively.
The contributions from $(jj)\oplus (j_bj_b)$ final state are smaller
by a factor of approximately 12, independently of the configuration
and transverse momentum thresholds, and assuming
$\epsilon_{fake}=0.01$.
  
Events  of the kind $pp\to b\bar{b}~\oplus~pp\to b\bar{b}$ turn out to
be totally negligible, at the level of 40 with the 
$p_{\rm T}^j>80$~GeV cut. 
 
The  events from three separate $pp$ collisions
contribute less than 10\% of the two-collision rates shown in the
Tables~\ref{tab:sb60a}-\ref{tab:sb80b}, at $10^{34}$~cm$^{-2}$s$^{-1}$. 

%%%%%%%%%%%%%%%%%%% III TABLE
\begin{table}[h]
\begin{center}
\ccaption{*}{\label{tab:sens60}
The sensitivity, defined as the ratio of the number of signal events 
divided by the square root of the number of the background events. 
The mistagging efficiency of 
light jets, $\epsilon_{fake}$, is $\epsilon_{fake}=0.01$.
The integrated 
luminosity is 600~fb$^{-1}$ for both configurations (a),(b), 
and the transverse momentum cut on jets is $p_{\rm T}^j > 60$~GeV.
}
\vskip 2mm
\begin{tabular}{||l|l|l|l||}\hline
$m_H$ & 115~GeV  & 120~GeV  & 140~GeV \\
\hline
$(a)~S/\sqrt{B}$ 
%& $3.3$  & $3.2$   & $1.5$ \\
                 & $3.0$  & $2.9$   & $1.4$ \\            
\hline
$(b)~S/\sqrt{B}$ 
%& $5.6$  & $5.4$   & $3.1$ \\
                & $5.1$   & $5.2$   & $2.7$ \\ 
\hline
\end{tabular}            
\end{center}
\end{table}

%%%%%%%%%%%%%%%%%%% III TABLE pt80
\begin{table}[h]
\begin{center}
\ccaption{*}{\label{tab:sens80}
The same as Table~\ref{tab:sens60}, with $p_{\rm T}^j > 80$~GeV.}
\vskip 2mm
\begin{tabular}{||l|l|l|l||}\hline
$m_H$ & 115~GeV  & 120~GeV  & 140~GeV \\
\hline
$(a)~S/\sqrt{B}$ 
%& $2.4$  & $2.3$   & $1.1$ \\
                 & $2.4$  & $2.3$   & $1.0$ \\            
\hline
$(b)~S/\sqrt{B}$ 
%& $4.1$  & $4.1$   & $2.3$ \\
                 & $3.7$  & $4.1$   & $2.0$ \\ 
\hline
\end{tabular}            
\end{center}
\end{table}

%%%%%%%%%%%%%%%%%%% IV TABLE
\begin{table}[h]
\begin{center}
\ccaption{*}{\label{tab:sens605}
The same as Table~\ref{tab:sens60} but with a mistagging 
efficiency of $\epsilon_{fake}=0.05$. }
\vskip 2mm
\begin{tabular}{||l|l|l|l||}\hline
$m_H$ & 115~GeV  & 120~GeV  & 140~GeV \\
\hline
$(a)~S/\sqrt{B}$ 
%& $2.7$  & $2.7$   & $1.2$ \\
                 & $2.5$  & $2.4$   & $1.1$ \\
\hline
$(b)~S/\sqrt{B}$ 
%& $4.5$  & $4.4$   & $2.4$ \\
                 & $4.4$  & $4.2$   & $2.1$ \\ 
\hline
\end{tabular}            
\end{center}
\end{table}

%%%%%%%%%%%%%%%%%%% IV TABLE PT>80
\begin{table}[h]
\begin{center}
\ccaption{*}{ \label{tab:sens805}
The same as Table~\ref{tab:sens80} but with a mistagging 
efficiency of $\epsilon_{fake}=0.05$. }
\vskip 2mm
\begin{tabular}{||l|l|l|l||}\hline
$m_H$ & 115~GeV  & 120~GeV  & 140~GeV \\
\hline
$(a)~S/\sqrt{B}$ 
%& $2.2$  & $2.2$   & $1.0$ \\
                 & $2.2$  & $2.1$   & $1.0$ \\
\hline
$(b)~S/\sqrt{B}$ 
%& $3.5$  & $3.5$   & $1.9$ \\
                 & $3.1$  & $3.3$   & $1.6$ \\ 
\hline
\end{tabular}            
\end{center}
\end{table}

%%%%%%%%%%%%%%%%%%% V TABLE
\begin{table}[h]
\begin{center}
\ccaption{*}{\label{tab:ssig60}
The statistical significance of the determination of 
the branching ratio $\Gamma_b / \Gamma$ and of the
$b$-quark Yukawa coupling 
in the configurations (a) and (b).
A luminosity of $600$~fb$^{-1}$ is assumed; 
the transverse momentum cut on jets 
is $p_{\rm T}^j > 60$~GeV. Here $\epsilon_{fake}=0.01$.
Using $\epsilon_{fake}=0.05$ will worsen these estimates by
approximately 20\%.}
\vskip 2mm
\begin{tabular}{||l|l|l|l|l||}\hline
& $m_H$ & 115~GeV  & 120~GeV  & 140~GeV \\
\hline
$(a)$ & $\delta \Gamma_b/\Gamma$ & $0.33$  & $0.35$   & $0.71$ \\
 & $\delta y_{Hbb}/y_{Hbb}$ & $0.58$  & $0.51$   & $0.56$ \\
\hline
$(b)$ & $\delta \Gamma_b/\Gamma$ & $0.20$  & $0.19$   & $0.37$ \\ 
 & $\delta y_{Hbb}/y_{Hbb}$ & $0.36$  & $0.30$   & $0.29$ \\ 
\hline
\end{tabular}            
\end{center}
\end{table}

%%%%%%%%%%%%%%%%%%% V TABLE pt > 80
\begin{table}[h]
\begin{center}
\ccaption{*}{\label{tab:ssig80}
The same as Table~\ref{tab:ssig60} with  $p_{\rm T}^j > 80$~GeV.}
\vskip 2mm
\begin{tabular}{||l|l|l|l|l||}\hline
& $m_H$ & 115~GeV  & 120~GeV  & 140~GeV \\
\hline
$(a)$ & $\delta \Gamma_b/\Gamma$ & $0.42$  & $0.43$   & $1$ \\
 & $\delta y_{Hbb}/y_{Hbb}$ & $0.76$  & $0.68$   & $0.72$ \\
\hline
$(b)$ & $\delta \Gamma_b/\Gamma$ & $0.27$  & $0.24$   & $0.50$ \\ 
 & $\delta y_{Hbb}/y_{Hbb}$ & $0.47$  & $0.40$   & $0.36$ \\ 
\hline
\end{tabular}            
\end{center}
\end{table}

\section{Results}
Tables~\ref{tab:sens60}-\ref{tab:sens805} summarize our results for 
the sensitivity defined as the ratio of the number of signal events 
divided by the square root of the number of background events 
for different values of the mistagging 
efficiency $\epsilon_{fake}$. 
Tables~\ref{tab:ssig60},\ref{tab:ssig80} show our results on 
the determination of the  branching ratio ${\cal B}(H \to b \bar{b})$
and accordingly on the $H b\bar{b}$ Yukawa coupling 
$y_{Hbb}$, assuming the knowledge of the $HWW$
coupling. This can be determined using other channels, as discussed
in the literature~\cite{HWW}. These results rely also on the
assumption of $SU(2)$ invariance to relate the contributions to the
signal coming from the $HWW$ and $HZZ$ couplings, which can not be
experimentally disentangled in the WBF production mechanism. 
With a
total luminosity of 600~fb$^{-1}$, a relative precision of about 20\%
on the ${\cal B}(H\to b\bar{b})$ branching ratio can be
attained.  This represents an improvement with respect to what
obtained in other channels~\cite{zeppcoupl,belyaevreina}.  As for the
$Hb\bar{b}$ Yukawa coupling, a statistical significance of at best  
$30\%$ is reachable~\footnote{The statistical significance of 
the $b$-quark Yukawa coupling is linked to the one of the branching 
ratio by the following formula: $\delta y_{Hbb}/y_{Hbb} = 
\delta {\cal B} / (2 {\cal B} ({1 - \cal B}))$, where ${\cal B}$ stands for 
the branching ratio $H \to b \bar b$.}. The significance is rather
flat in the 115-140~GeV mass range, as a result of the compensation
between overall rate (which decreases at larger masses) and
sensitivity of the BR to the Yukawa coupling (sensitivity which
increases at smaller BR, for larger masses). The effect of applying
a larger cut (80~GeV) on the transverse momentum of forward jets is to
reduce by approximately 10\% the statistical accuracy of the
measurement. This choice could however turn out to be more reasonable
in view of the reduced experimental difficulties at larger $p_{\rm T}^j$.

The $H\to b\bar{b}$ decay in the WBF channel also allows for a 
model independent determination of the ratio of 
widths $\Gamma(H\to b\bar{b})/\Gamma(H\to \tau^+\tau^-)$ when combined
with the $qq\to qq(H\to \tau^+\tau^-)$ mode~\cite{zepp2}. 
This determination can be compared with what obtained in the $t\bar{t}H$
production channel by~\cite{belyaevreina}.
Moreover, comparing the WBF mechanism studied in this paper with 
the associated $W(H\to b\bar{b})$ production, 
one could test the $SU(2)$ relation between the SM $HWW$ and $HZZ$ couplings
for low Higgs masses.

\section{Conclusions}
In this letter we examined 
 $(H\to b\bar{b})jj$ production at the LHC, with the goal
of assessing the potential accuracy in the  determination of 
the $y_{Hbb}$ Yukawa coupling.
A study of the observability of this channel has also been presented in
 ref.~\cite{DeRoeck:2002hk}. We believe our paper provides a more
 realistic evaluation of the experimental challenges of this
 measurement, and find less optimistic results.

In particular, we identified two main sources of  backgrounds:
\begin{itemize}
\item 4 jet final states: these are over 100 times larger than the
signal, but could be  evaluated with accuracy using the sidebands of
the $b\bar{b}$ mass spectrum. This requires however some tagging
information to be available at the trigger level, to reduce to
acceptable levels the data storage needs for inclusive, untagged, 4
jet final states.
\item 4 jet final states from multiple collisions: a large
contribution comes from events of the type $(jj_b) \oplus (jj_b)$,
where the $b\bar{b}$ mass spectrum has a broad peak in the middle of
the signal region. The absolute rate of these events (of the order of
the signal rate, when using the lower transverse momentum threshold of
60~GeV) can be determined if the distribution of the $z$ vertex separation
between the two overlapping events can be determined with a resolution of
the order of 5-10mm. These events are significantly reduced in number
when using the higher threshold of 80~GeV for the forward jets.
\end{itemize}
Our parton-level analysis should be completed with a full detector
simulation, but, already at this stage, it provides a strong
indication for the relevance of this channel for the ${\cal B}(H\to
b\bar{b})$ branching ratio.  We have shown in fact that the ${\cal
B}(H\to b\bar{b})$ can be measured with a $20\%$ precision for an
Higgs mass around $120$~GeV assuming that the coupling $HWW$ is the
one predicted by the Standard Model or determined in other reactions
already studied in the literature.  We also observe that the WBF
channel we study, combined with other processes, can be used for a
model independent determination of the $y_{Hbb}/y_{H\tau\tau}$ ratio
and for a test of the ratio of the couplings $g_{HWW}/g_{ZWW}$ for low
Higgs masses.

To conclude, we should point out that all statistical accuracies
listed in this study should be matched by an excellent control over
experimental systematics, including the knowledge of $b$-tagging
efficiencies (needed for example to allow the determination of $Z\to
b\bar{b}$ backgrounds from the measurement of $Z\to \ell^+\ell^-$
final states) and their dependence on the $b$ momentum, and of forward
jet tagging efficiencies and fake (pile-up or calorimeter noise) rates. 
On the other hand, as
mentioned at the beginning, we expect our estimates of the physics
backgrounds to be very conservative, being based on very low $Q^2$
scales for the evaluation of the strong coupling constant;
furthermore, we anticipate that more sophisticated analyses based on
kinematical correlations in the event (exploiting for example the
scalar nature of the $Hb\bar{b}$ coupling) will help improving the
signal significance. 
\newpage
{\bf Acknowledgements} \par We wish to thank A. Djouadi, F. Gianotti,
K. Jakobs and G. Polesello for useful discussions. FP thanks the Pavia
Gruppo IV of INFN for access to the local computing resources.
We thank in particular A. Djouadi for pointing out a silly mistake in
our evaluation of the Yukawa coupling significance in the first
version of this work.


\begin{thebibliography}{10}

\bibitem{cmsnote} V. Drollinger, T. M\"uller and D. Denegri,
hep-ph/0111312; hep-ph/0201249.

\bibitem{cmsnote2} E. Richter-Was, Acta. Phys. Pol {\bf B30}, 
1001 (1999); ibid. {\bf B31}, 1973 (2000).

\bibitem{duca} V. Del Duca, W. Kilgore, C. Oleari, C. Schmidt 
and D. Zeppenfeld, Phys. Rev. Lett. {\bf 87}, 122001 (2001).

\bibitem{zepp} D. Rainwater, D. Zeppenfeld and K. Hagiwara, Phys. Rev.
{\bf D59}, 014037 (1999).

\bibitem{DeRoeck:2002hk}
A.~De Roeck, V.A.~Khoze, A.D.~Martin, R.~Orava and M.G.~Ryskin,
%``Ways to detect a light Higgs boson at the LHC,''
Eur.\ Phys.\ J.\  {\bf C25} (2002) 391
[arXiv:hep-ph/0207042]; V.A.~Khoze, M.G.~Ryskin, 
W.J.~Stirling, P.H.~Williams, hep-ph/0207365.
%%CITATION = HEP-PH 0207042;%%

\bibitem{alpgen} M.L. Mangano, M. Moretti, F. Piccinini, R. Pittau
and A.D. Polosa, hep-ph/0206293; 
M.L.~Mangano, M.~Moretti and R.~Pittau, 
Nucl. Phys. {\bf B632}, (2002) 343. 

\bibitem{alpha}
F.~Caravaglios and M.~Moretti, \plb{358}{1995}{332}; 
F. Caravaglios, M.L. Mangano, M. Moretti and R. Pittau, 
Nucl. Phys. {\bf B539}, 215 (1999).

\bibitem{Alitti:1990kw}
J.~Alitti {\it et al.}  [UA2 Collaboration],
Z.\ Phys.\  {\bf C49} (1991) 17.

\bibitem{HWW}  
D.~Rainwater and D.~Zeppenfeld, Phys. Rev. {\bf D60}, 113004 (1999);
N.~Kauer, T.~Plehn, D.~Rainwater, and D.~Zeppenfeld, 
Phys. Lett. {\bf B503}, 113 (2001).

\bibitem{zeppcoupl} D.~Zeppenfeld, hep-ph/0203123.

\bibitem{belyaevreina} A.~Belyaev and L.~Reina, hep-ph/0205270.

\bibitem{zepp2} D.~Zeppenfeld, R.~Kinnunen, A.~Nikitenko and
E.~Richter-Was, Phys. Rev. {\bf D62}, 013009 (2000). 





\end{thebibliography}
\end{document}